\documentclass[twocolumn,showpacs,floatfix,amsmath,amssymb,prb,superscriptaddress]{revtex4}

\usepackage{graphicx}
\usepackage{subfigure}
\makeatletter

\begin{document} 
\newcommand{\be}{\begin{equation}}
\newcommand{\ee}{  \end{equation}}
\newcommand{\ba}{\begin{eqnarray}}
\newcommand{\ea}{  \end{eqnarray}}
\newcommand{\ve}{\varepsilon}

\title{Mesoscopic features in the transport properties of a Kondo-correlated quantum dot in a magnetic field}

\author{Alberto Camjayi} \affiliation{Departamento de F\a'{i}sica,
  FCEyN and IFIBA, Universidad de Buenos Aires, Pabell\'on 1, Ciudad
  Universitaria, 1428 Buenos Aires, Argentina}

\author{Liliana Arrachea} \affiliation{Departamento de F\a'{i}sica,
  FCEyN and IFIBA, Universidad de Buenos Aires, Pabell\'on 1, Ciudad
  Universitaria, 1428 Buenos Aires, Argentina}

\date{\today}

\begin{abstract}
We study the transport behavior induced by a small bias voltage through a quantum dot connected to 
one-channel finite-size wires. We describe the quantum dot by the Hubbard-Kondo which is solved by means of a quantum Monte Carlo method.
We investigate the effect of a magnetic field applied at the quantum dot in the Kondo regime. We identify changes in the behavior of 
mesoscopic oscillations introduced by the magnetic field that have an analogous behavior to those observed as a function of the temperature.
\end{abstract}


\maketitle

\section{Introduction}
The consequences of the Kondo effect on the transport properties of mesoscopic systems and nanostructures 
 has received a significant interest during the past two decades. After signatures of this effect were identified in quantum dots 
fabricated in semiconducting systems,\cite{gold-gor} it was later investigated in several other devices, containing carbon nanotubes, 
and molecules attached to metallic electrodes.\cite{kondomes,rev,kondonan}

All these systems contain a central piece, the quantum dot, which is connected to wires of non-interacting electrons. The Kondo effect 
takes place as a consequence of the Coulomb repulsion, which originates an effective coupling between the spin of a localized electron 
in the quantum dot with the spin of the electrons of the wires. For temperatures lower than the Kondo temperature $T_K$, these 
electrons form a singlet giving rise to a resonant state,  which manifests itself as the opening of a transport channel for each spin 
component, corresponding to a conductance $G_0=2 e^2/h$.\cite{hewson} The electrons of the wires that intervene in the formation of 
these singlets define the so called ``screening cloud.'' The latter extends up to a length that is related to the 
Kondo temperature as $\xi_K \sim \hbar v_F /k_B T_K$, being $v_F$ the Fermi velocity of the
electrons of the wire. The nature of the wires is known to play a crucial role in the 
development of the Kondo effect and the concomitant behavior of $G$. In mesoscopic systems and nanostructures, where the
wires have a finite length $L$ and a typical level spacing $\Delta$, the ratio 
$\xi_K/L$ (or, equivalently $\Delta/T_K$) defines different regimes for the behavior
of the Kondo effect and $G$. This effect has been analyzed in detail for the case of clean wires in Refs.~\onlinecite{sim-af, corn-bal}.
When the mesoscopic wires are dirty, there is an interesting interplay between the length of the Kondo cloud and the localization length introduced by disorder, which has been recently analyzed.\cite{us}

Another interesting aspect is the influence of other magnetic phenomena in the formation of the Kondo state. This issue has been
analyzed in the context of a Kondo impurity connected to a ferromagnetic environment,\cite{14} and for a quantum dot in an external 
magnetic field.\cite{costi, rusos, rosa,17,18,smir-grif}
The formation of singlets leading to the development of Kondo resonance, is clearly affected by the presence of an external 
magnetic field. In particular, the Zeeman splitting of the energy levels of the quantum dot destroys the Kondo resonance for strong 
enough magnetic fields. Interestingly, the conductance preserves the universality  under the influence of the magnetic field, which 
means that the Kondo temperature $T_K$ sets the energy scale for the behavior of this quantity.\cite{hewson} This property has been 
recently investigated experimentally\cite{18} and theoretically.\cite{smir-grif}  All these analysis have been performed for the case 
of quantum dots with a perfect matching to the wires. They find that the thermal and the magnetic energy, scaled  by the Kondo energy 
$k_B T_K$, play a similar role in the behavior of the conductance.
The aim of the present work is to analyze if this is also the case for a quantum dot embedded in mesoscopic wires with a finite length.

\begin{figure}[tb]
\includegraphics[height=6cm,keepaspectratio=true]{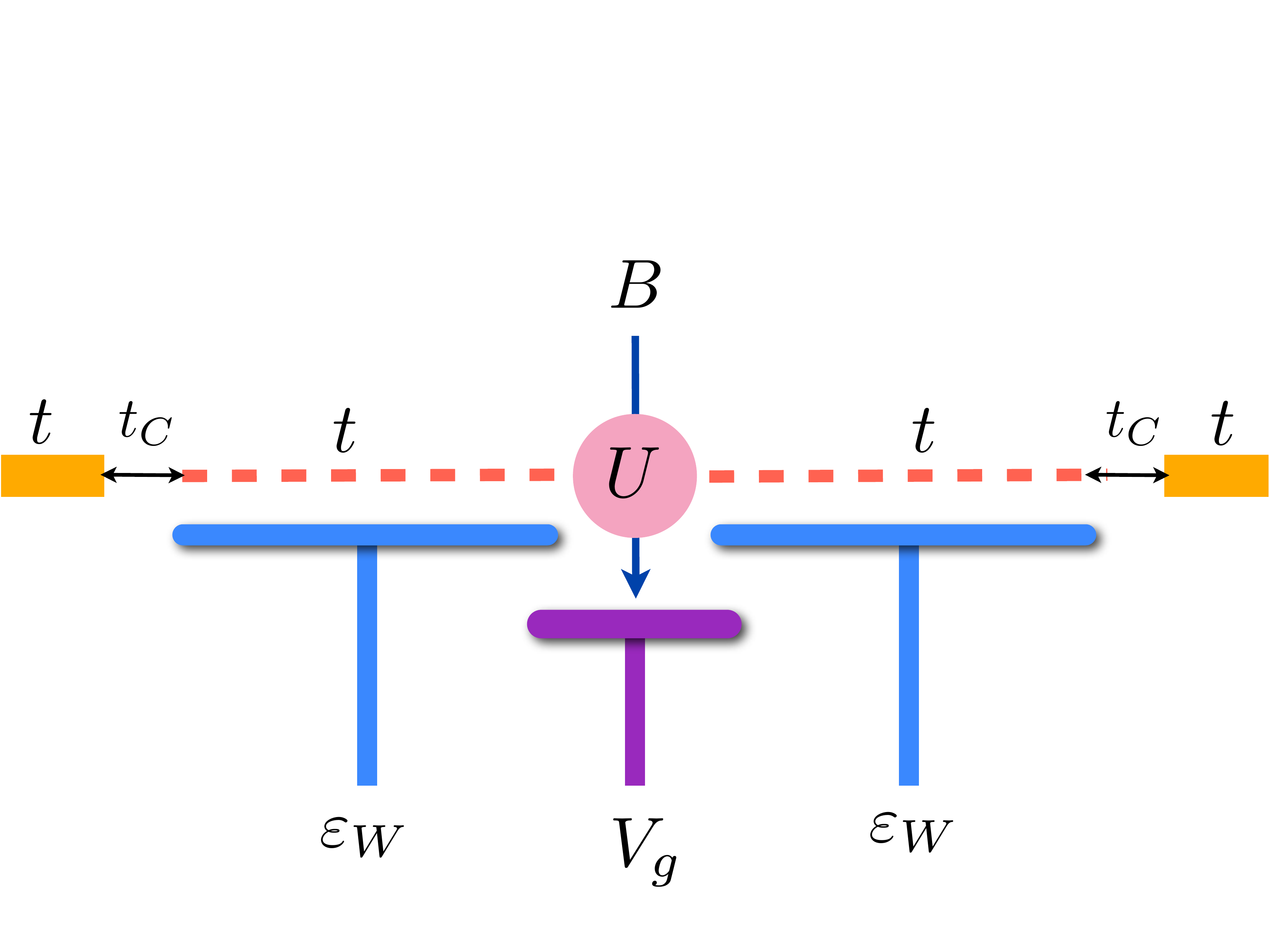}
 \caption{(Color online) Sketch of the considered setup.  An interacting quantum dot with Coulomb repulsion $U$ a local gate voltage
$V_g$ and a local magnetic field $B$ is connected to left and right finite-size wires. The latter have local are connected to 
macroscopic reservoirs through a hopping element $t_{C}$.}
\label{fig0}
\end{figure}

 We model the dot by an Anderson impurity. An important side of the study of the Kondo effect in this model is to tackle it with a 
reliable many-body technique. In this sense, treatments based on Bethe Ansatz,\cite{bethe} slave bosons,\cite{slabos} perturbation 
theory in different limits,\cite{pertheor} renormalization group,\cite{nrg,costi,rosa} exact diagonalization\cite{exdia} and quantum 
Monte Carlo methods,\cite{qmc} have been widely used in the literature.
In the context of transport properties, quantum Monte Carlo is, perhaps the less explored technique, except for a few examples.\cite{lilichac,egger,us} We show that it is indeed a very powerful one. In particular, we recover all the results presented for the 
conductance of a Kondo-correlated quantum dot embedded in mesoscopic wires\cite{sim-af, corn-bal} as well as those for a magnetic field 
and wires with perfect matching.\cite{18,smir-grif}  We then present results for finite-size wires and magnetic field and show that, as 
in the case of perfect matching, the magnetic energy plays a similar role as the thermal energy in the behavior of the conductance. 
The paper is organized as follows. In Section II we present the model, the expression for the conductance in terms of Green functions 
and the quantum Monte Carlo procedure followed to evaluate these functions. In Section III we present and discuss the results without 
and with magnetic field. Section IV is devoted to summary and conclusions.

\section{Theoretical Approach}

\subsection{Model} \label{model}
The full system is sketched in Fig.~\ref{fig0}. It consists in a quantum dot modeled by an Hubbard-Anderson impurity
connected to one-dimensional left ($L$) and right ($R$) wires. A magnetic field is applied at the dot.
The ensuing Hamiltonian is
\be
H=H_{d} + \sum_{\alpha=L,R} H_{\alpha}+ H_{cont},
\ee
where the Hamiltonian for the dot includes the effect of the Coulomb repulsion $U$, the voltage gate $V_g$ and the Zeeman splitting
$\Delta_z= g \mu_B B$ due to the external magnetic field $B$. We assume $B$ applied along the $z$-direction, $\mu_B$ is the Bohr 
magneton and $g$ is the gyromagnetic factor. It reads
\be
H_{d} =   \sum_{\sigma= \uparrow, \downarrow} \varepsilon_{d,\sigma} n_{d,\sigma} + U n_{d,\uparrow} n_{d, \downarrow},
\ee
where $\varepsilon_{d,\uparrow \downarrow}= V_g \pm \Delta_z/2$. We distinguish a finite region with $N_{\alpha}$ sites of the wires, 
between the dot and the macroscopic reservoirs. Following previous works,\cite{sim-af,corn-bal} we model this piece by a tight-binding Hamiltonian with nearest neighbors hopping element $t$. We also consider a gate voltage represented by a uniform local energy $\ve _W$. 
The reservoirs are represented by semi-infinite tight-binding chains with hopping elements $t$. 
\begin{widetext}
\be
H_{ \alpha}  =   - t \sum_{l=1}^{N_{\alpha}-1}[ c^{\dagger}_{l,{\alpha},\sigma} c_{l+1,{\alpha},\sigma} + H. c. ] 
+ \ve_{W} \sum_{l=1, \sigma}^{N_{\alpha} -1}  c^{\dagger}_{l,{\alpha},\sigma} c_{l,{\alpha},\sigma}- t_C [ c^{\dagger}_{N_{\alpha},\sigma} c_{N_{\alpha}+1,\sigma} + H. c. ]
- t \sum_{l=N_{\alpha}}^{\infty}[ c^{\dagger}_{l,{\alpha},\sigma} c_{l+1,{\alpha},\sigma} + H. c. ].
\ee
\end{widetext}
The contacts between the wires and the quantum dot are described by the Hamiltonian
\be
H_{cont}=- t \sum_{\alpha=L,R}  [c^{\dagger}_{1,\alpha,\sigma} d_{\sigma} + H. c.].
\ee

In the limit of vanishing magnetic field, $\Delta_z=0$, the model reduces to 
the  Kondo impurity with an effective exchange constant $J= 2 t^2[1/(U+V_g)-1/V_g]$,\cite{hewson} which
displays a resonance at the Fermi level below the Kondo temperature defined from
\be \label{TK}
\frac{2}{J}=\int_{-\infty}^{+\infty}
\frac{d \omega}{\omega} \mbox{tanh}(\frac{\omega}{2 T_K}) \rho_{0}(\omega),
\ee
where $\rho_{0}(\omega)= -\mbox{Im}\left[\sum_{\alpha} g_{\alpha}^R(\omega) \right]$ is the density of states of the wires. The Kondo 
resonance is a consequence of the formation of a spin singlet state between the impurity and the conduction electrons at the wires,
which extends within a length $\xi_K= \hbar v_F/(k_B T_K)$. A mismatching 
in the contact between the wires and the reservoirs, $t_{C}  \neq t $  introduces finite-size 
features in $\rho_0(\omega)$, thus affecting the value of the Kondo temperature and the Kondo cloud, as discussed in 
Ref.~\onlinecite{sim-af,corn-bal} and the next subsection.

\subsection{Conductance}\label{conductance}
The conductance through the dot for an infinitesimal applied voltage $V$, such that the chemical potentials of the
$L$ ($R$) reservoirs are $\mu_{L,R}= \mu \pm V/2$ is:\cite{meiwi}
\ba \label{cond}
G& = & G_{\uparrow}+ G_{\downarrow}, \nonumber \\
G_{\sigma} & = & \frac{e^2}{\hbar} \int \frac{d \omega}{2 \pi} \frac{ \partial f (\omega)}{\partial \omega}
\frac{\Gamma_{L}(\omega) \Gamma_{R}(\omega)}{\Gamma_{L}(\omega) + \Gamma_{R}(\omega)}
\rho_{\sigma} (\omega).
\ea
The hybridization function with the wire 
\be
\Gamma_{\alpha}(\omega)  =   t^2 \rho_{\alpha}^0(\omega), \;\;\;
\rho_{\sigma} (\omega)  =  -2 \mbox{Im}[G^R_{d,\sigma}(\omega)],
\ee
depends on the coupling $t$ between the wire and the dot, and the density of states of the local density of states of the wire 
$\rho_{\alpha}^0(\omega)=- 2 \mbox{Im}[g^R_{\alpha}(\omega)]$, where $g^R_{\alpha}(\omega)$ is the retarded Green function of the wire  
$\alpha$ connected to the corresponding reservoir but uncoupled from the dot.
$G^R_{d,\sigma}(\omega)$ is the retarded Green function of the quantum dot connected to the wires. The Fermi function 
$f(\omega)=1/(1+e^{\beta (\omega-\mu)})$ depends on the temperature $T=1/(\beta k_B)$, as well as on the mean chemical potential of the 
reservoirs $\mu$. At $T=0$, the conductance simply results
\be \label{condt0}
G_{\sigma}(T=0)= \frac{e^2}{h} \frac{\Gamma_{L,\sigma}(\mu) \Gamma_{R,\sigma}(\mu)}{\Gamma_{L,\sigma}(\mu) + 
\Gamma_{R,\sigma}(\mu)}
\rho_{\sigma} (\mu) ,
\ee

While the calculation of the non-interacting Green functions entering $\Gamma_{\alpha}(\omega)$ is a very simple task (see appendix A), 
the calculation of the Green function $G^R_{d,\sigma}(\omega)$ implies using a suitable many-body method to treat the Coulomb 
interaction. In the next subsection, we present a procedure based on quantum Monte Carlo.

\subsection{Monte Carlo Method}
In the study of local properties of the Anderson impurity model, like the Green function $G^R_{d,\sigma}(\omega)$ 
quantum Monte Carlo (QMC) is one of the most efficient methods.  It allows for the numerical evaluation of  the Matsubara Green function 
${\cal G}_{d,\sigma}(i \omega_n)$, where $\omega_n = (2 n+1) \pi/ \beta$, with high a precision, since it is  free from the so called 
``sign'' problem which usually plagues QMC method in other fermionic models.
 In fact, this method
has been widely used in dynamical mean field theory  calculations,\cite{dmft} which involve a self-consistent evaluation of the Green 
function of the Anderson impurity model. In the context of quantum transport, this method has also been extended to treat
vibrational degrees of freedom.\cite{lilichac} 

From the first implementation of QMC due to Hirsch and Fye,\cite{hirfy} a new class of impurity solvers has been developed, the so 
called continuous time quantum Monte Carlo (CTQMC),\cite{review_gull,werner} which  set the basis
for  band structure calculations by using for example LDA+DMFT.\cite{elec-struc}  
In a previous work,\cite{us} we have used the CTQMC based on hybridization expansion\cite{haule} to investigate the conductance of a 
quantum dot in the Kondo regime connected to disordered wires and we follow a similar strategy here.

The non-interacting Green functions $g_{\alpha}(i \omega_n)$
 enter as input for the CTQMC to obtain ${\cal G}_{d,\sigma}(i \omega_n)$.
In order to evaluate $G_{\sigma}$ from (\ref{cond}) we also need the corresponding retarded Green functions $g^R_{\alpha}(\omega)$. 
The functions $g_{\alpha}(i \omega_n)$ and $g^R_{\alpha}(\omega)$ are evaluated by using the
recursive procedure of Appendix A. The delicate step consists in the analytical continuation of the output functions of the
Monte Carlo method, ${\cal G}_{d,\sigma}(i \omega_n)$, to the real frequency axis, in order to obtain $G^R_{d,\sigma}(\omega)$.
A possible route to follow is a maximum entropy method.\cite{maxent} These methods proved to be efficient to capture relevant features 
of the spectral functions, like the number of peaks, as well as their positions, irrespectively of the magnitude of the
energies involved. The shortcoming is that they fail to accurately predict the corresponding width of the peaks. 
In the present work, we follow an alternative route, which consists in a low-frequency polynomial fit of the Green function on the 
imaginary frequency axis. This procedure is motivated in the fact that in Eq. (\ref{cond}) the derivative of the Fermi function defines 
for low temperatures a very narrow integration window of energies around the mean chemical potential of the wires $\mu$. For practical 
purposes, we set $\mu=0$ and change the voltage gate $V_g$. Thus, in order to compute the conductance
all we need is a reliable analytical continuation within this narrow interval of energies centered at $\omega=0$. From the
 CTQMC simulation we can obtain not only the Green function but also the self-energy in the Matsubara axis
 $\Sigma(i \omega_n)=  {\cal G}_{d,\sigma}(i \omega_n)^{-1} -{\cal G}^0_{d,\sigma}(i \omega_n)^{-1} $. In the metallic regime, 
within which the Kondo effect takes place, this function is known to behave as a Fermi liquid, for which $\mbox{Im}[\Sigma^R( \omega)]
\propto \omega ^2$ close $\omega=0$. It is, then, natural to expect a behavior for  $\Sigma(i \omega_n)$ that can be fitted by a 
quadratic function of $i \omega_n$. In our case, we keep the discrete CTQMC values for 
$\Sigma(i \omega_n)=\Sigma^{\prime}(i \omega_n) + i \Sigma^{\prime \prime}(i \omega_n),\;n=0, \ldots N$, with $N \sim 10$ and
fit  the functions $\Sigma^{\prime, (\prime \prime)}(i \omega_n)$  by functions of the form
$y (i \omega) = a^{\prime, (\prime \prime)}+ b^{\prime, (\prime \prime)} (i\omega) +  c^{\prime, (\prime \prime)} (i\omega)^2 $. 
This procedure can be carried out with a very high accuracy and the analytical continuation can be easily done by simply substituting 
$i \omega \rightarrow \omega + i \eta$, with $\eta= 0^+$ in the  function $y(i \omega)$.

\section{Results}
\subsection{Without magnetic field}\label{sin_campo}
\begin{figure}[htb]
\centering
\includegraphics[clip,width=1\linewidth]{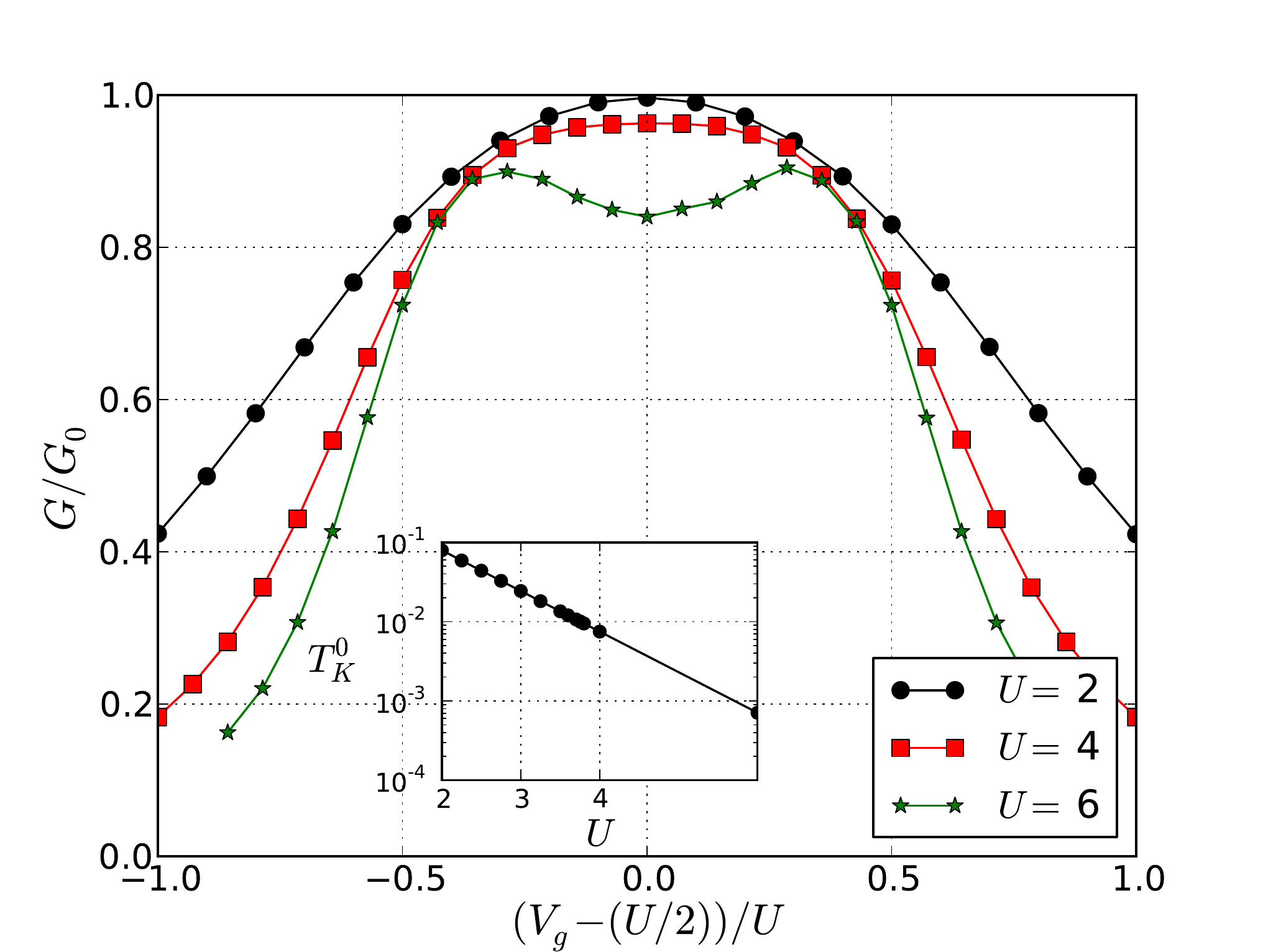}
\caption{\label{clean}Balistic wires. Conductance for several values of $U$ for a fixed value of the temperature
$T=0.01$. Inset: semi-log plot of $T_K^0$ vs. $U$.}
\end{figure}

A relevant test for the numerical procedure based on quantum Monte Carlo is the analysis of the conductance 
$G$ of the interacting quantum dot as a function of the gate voltage $V_g$. The typical behavior for this quantity
has been evaluated with several methods and published in several works.\cite{sim-af,corn-bal,pertheor,exdia}
As a benchmark, we have checked that  our method is able to recover the main features discussed in the literature.
Typical plots are shown in Fig.~\ref{clean} for three different values of the Coulomb interaction $U$ and $T=0.01$, and
wires, characterized by the same hopping parameter as that of the reservoirs, and perfect matching, i.e. $t_{C}= t$.
Each of the values of the Coulomb repulsion shown in the figure define different values of $J$ which, in turn, imply different Kondo temperatures $T_K^0$.
These can be evaluated from  Eq.~(\ref{TK}) and the results are shown in the inset of Fig.~\ref{clean}. Below the Kondo temperature
the Kondo resonance is developed opening a conduction channel. The conductance, thus, equals the conductance quantum $G_0$ for each 
spin degree of freedom, a situation which is referred to as the ``unitary limit.'' Typically this takes place with an interval of gate potentials $-U \leq V_g-U/2 \leq U $.

For $U = 2$, we show that the conductance 
corresponds to a situation between the so called mixed-valence regime and the Kondo regime. In this case, the unitary limit
$G = G_0$ is achieved only for $V_g = U/2$, displaying a peak with a width $\sim U$. 
For $U=4$, the temperature is  $T \sim T_K^0$ for that value of $U$ and we clearly see that $G$ displays a behavior very close to that of the
unitary limit, with a plateau with $G \sim G_0$ within
 $-U \leq V_g-U/2 \leq U $.
Finally, we show the case of $U = 6$,
which corresponds to a Kondo temperature $T_K^0 = 7\times10^{-4} t$, which is lower than the temperature considered in our simulations.
 In comparison to the behavior of $G$ 
in the other cases, we can see that a valley appears within the range of $V_g$ corresponding to the plateau in the Kondo regime, 
while two Coulomb blockade peaks centered at $V_g-U/2=\pm U/2$ develop.

\begin{figure}[htb]
\centering
\includegraphics[clip,width=1\linewidth]{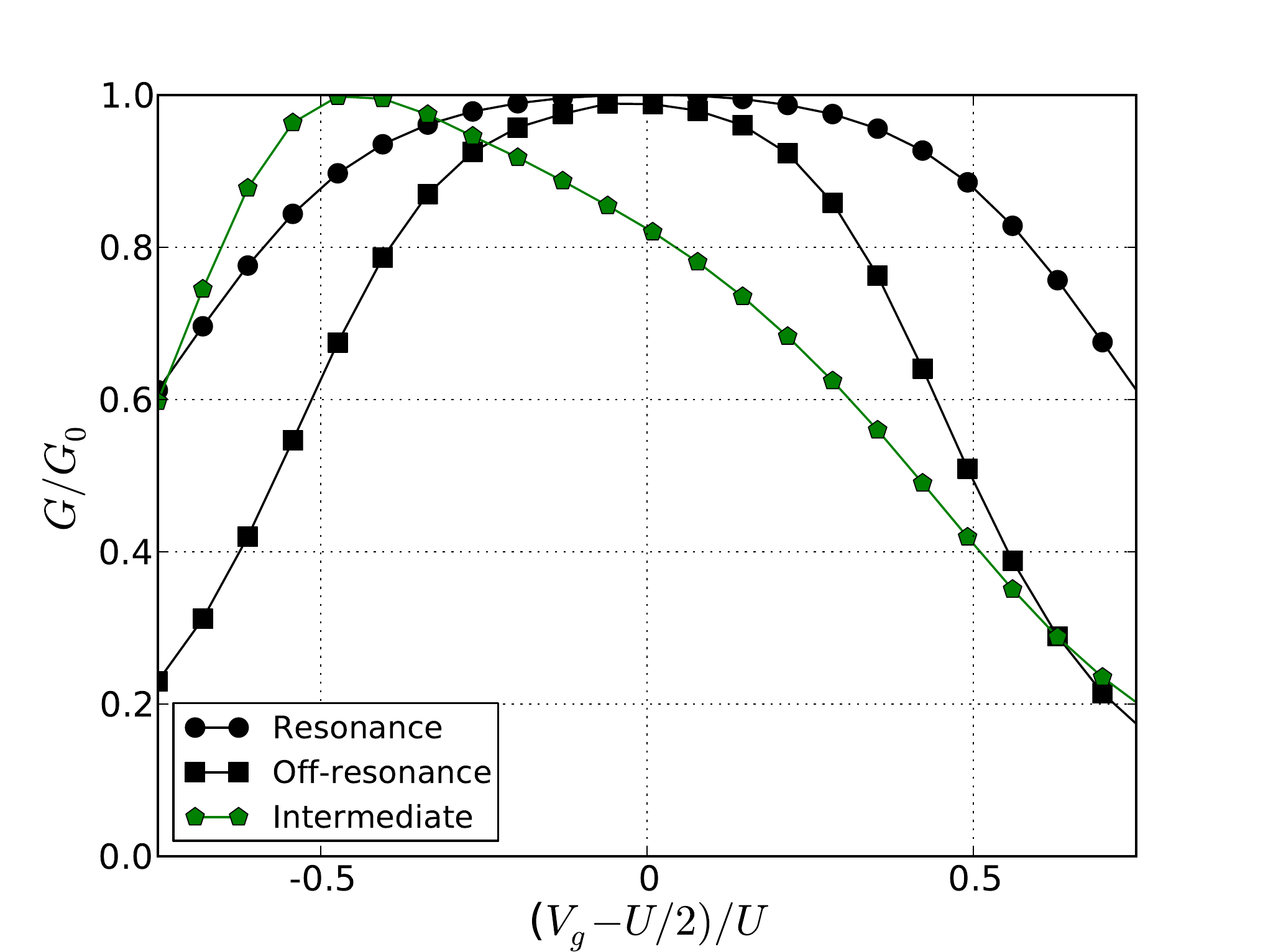}
\caption{Conductance as a function of gate voltage for finite wires with mismatching, $t_{C}=0.8t$.
The three curves correspond to different values of $\ve_W$, at resonance (circles), off-resonance (squares) and intermediate case (pentagons).
Other parameters are $V_g=0$ and  $T=0.0025$. $N_L=N_R=7$.} \label{gmist0}
\end{figure}
In order to verify that the quantum Monte Carlo method is also able to capture more subtle mesoscopic effects of the Kondo cloud,
we analyze the behavior of the conductance in a  wire with a mismatching in the contact modeled by $t_{C} <t$ and 
finite local energy within the finite-length wire $\ve_{W}\neq V_g$.  In this case, the wires and the dot define a finite-size box. The
spectral density of the wires corresponds to a sequence of peaks at the energy of the internal levels. 
We consider left and right chains of equal length, $N_L=N_R=N$.
The corresponding mean level spacing of the $N$-site chain  in this case is $\Delta \sim 4t/N$. In the Fig.~\ref{gmist0} we consider 
$\Delta= 0.4 t$, 
which satisfies $\Delta \gtrsim T_K^0$. In this regime, where the typical level spacing of the wires is concurrent with the Kondo 
temperature for ballistic wires, mesoscopic effects are expected to be relevant. \cite{sim-af, corn-bal} An equivalent picture for this 
scenario corresponds to a Kondo cloud of length $\xi_K < L$, being $L=Na$, where $a$ is the lattice constant of the wire.
In Fig.~\ref{gmist0} we show the low temperature behavior of the conductance as a function of the gate voltage for three different values
of $\ve_W$. Varying this parameter allows us to tune at the Fermi energy different regions of the energy spectrum of the wires. 
In one of the plots of the figure, we show the behavior of $G$ corresponding to parameters for which the 
Fermi energy of the infinite leads coincides with the energy of a resonant level of the finite wire (circles). In this case, the behavior of 
the conductance as a function of $V_g$ is basically the same as that of Fig.~\ref{clean}. A similar behavior is observed when the Fermi 
energy lies perfectly off-resonance (squares). This case corresponds to a valley between two consecutive peaks in the spectral function
of the wire. The density of states at the Fermi energy 
$\rho_0(\mu)$ is, thus, much smaller than in the resonant case and  Eq.~(\ref{TK}) leads to a Kondo temperature $T_K^* < T_K^0$. The 
data of the Fig.~\ref{gmist0} corresponds to a 
temperature $T< T_K^*$. For this reason, the unitary limit of the conductance $G \sim G^0$ can be observed for $V_g \sim 0$,  although
the conductance plateau as a function of $V_g$ is in this case narrower than for the resonant case.
We also illustrate an intermediate situation (pentagons), which does not exactly correspond nor to  resonance neither to off-resonance, where 
anomalous asymmetric features are observed. For sufficiently large chains with $\Delta \ll T_K^0$, all these finite-size effects become irrelevant 
and the behavior of Fig.~\ref{clean} is recovered. The picture for this case is a Kondo cloud fitting within the length of the wires.

\begin{figure}[htb]
\centering 
\includegraphics[clip,width=1\linewidth]{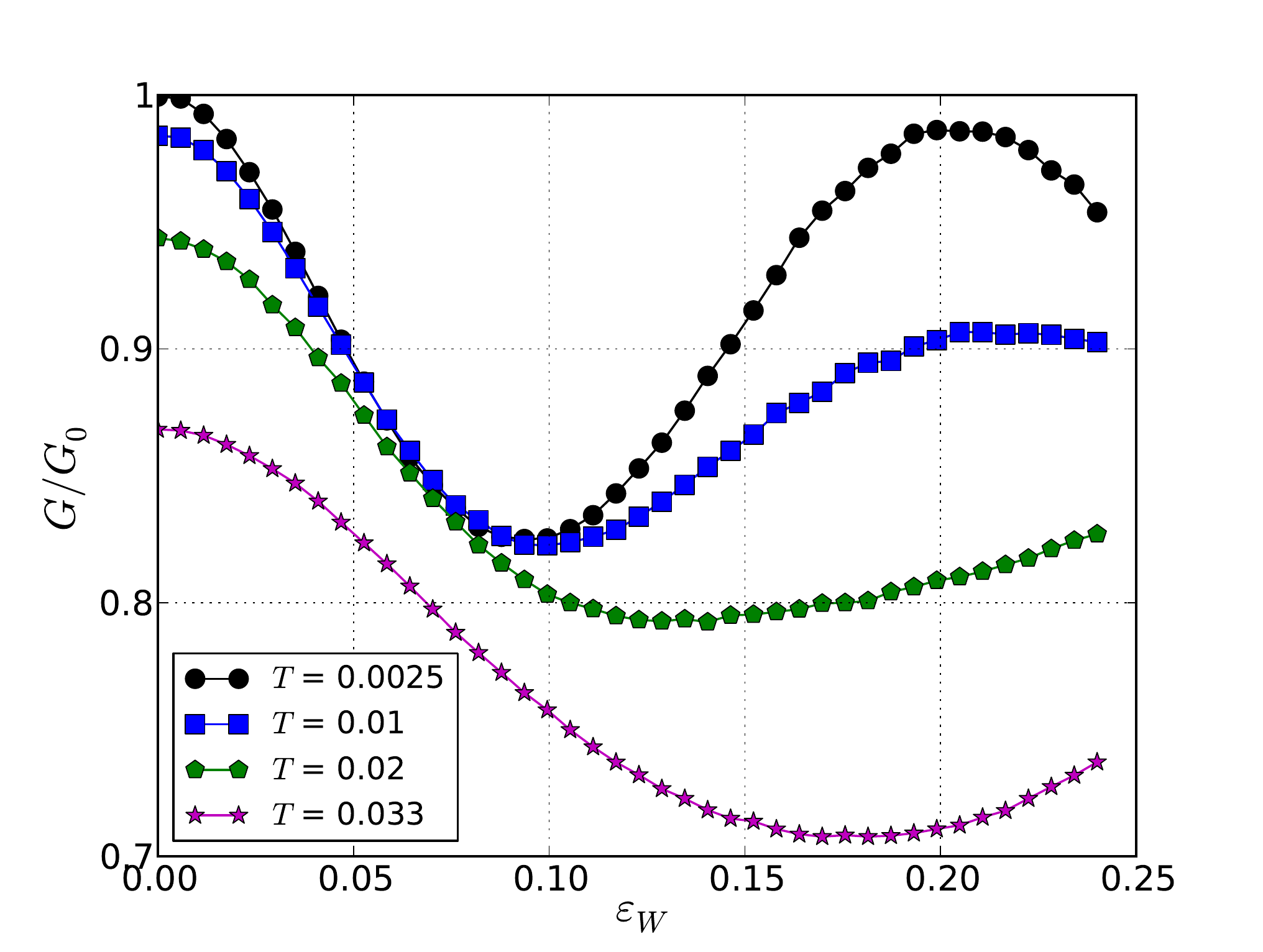}
\caption{Conductance as a function of the energy shift $\ve_W$ of the wires for finite wires with mismatching, $t_{C}=0.8t$.
Different plots correspond to different temperatures, $T<T_K^0$ (circles),  $T \sim T_K^0$ and $T>T_K^0$ (pentagons and stars). 
Other parameters are the same as in Fig.~\ref{gmist0}.}\label{figos0}
\end{figure}

In Fig.~\ref{figos0} we observe another interesting effect of the finite-size wires. Namely the oscillations of the conductance as a 
function of the energy $\ve_W$ of the wires and the change of the ensuing periodicity as the temperature grows above the Kondo 
temperature $T^*_K$. For these plots we keep the gate voltage of the dot fixed at $V_g=0$. The low temperature regime can be understood 
from the behavior discussed in Fig.~\ref{gmist0}, where we found maxima of the conductance $G= G^0$ for both resonant and off resonant configurations. Recalling that $\Delta$ is the mean level spacing of the wires, this corresponds to
 a periodicity equal to $\Delta/2$, consistent with a maximum of the conductance when the value of $\ve_W$ tunes a resonant as well as 
an off-resonant energy of the wire at the Fermi energy. The opposite limit for $T\gg T_K^0$, where the Kondo resonance is completely 
melted and the dot enters the Coulomb blockade regime, the conductance displays maxima only when $\ve_W$ tunes a resonant level of the 
wire at the Fermi energy. The change of the periodicity of the conductance as a function of $\ve_W$ as $T$ increases is clearly 
observed in the figure.

\subsection{Effect of the magnetic field}
\begin{figure}[htb]
\centering 
\includegraphics[clip,width=1\linewidth]{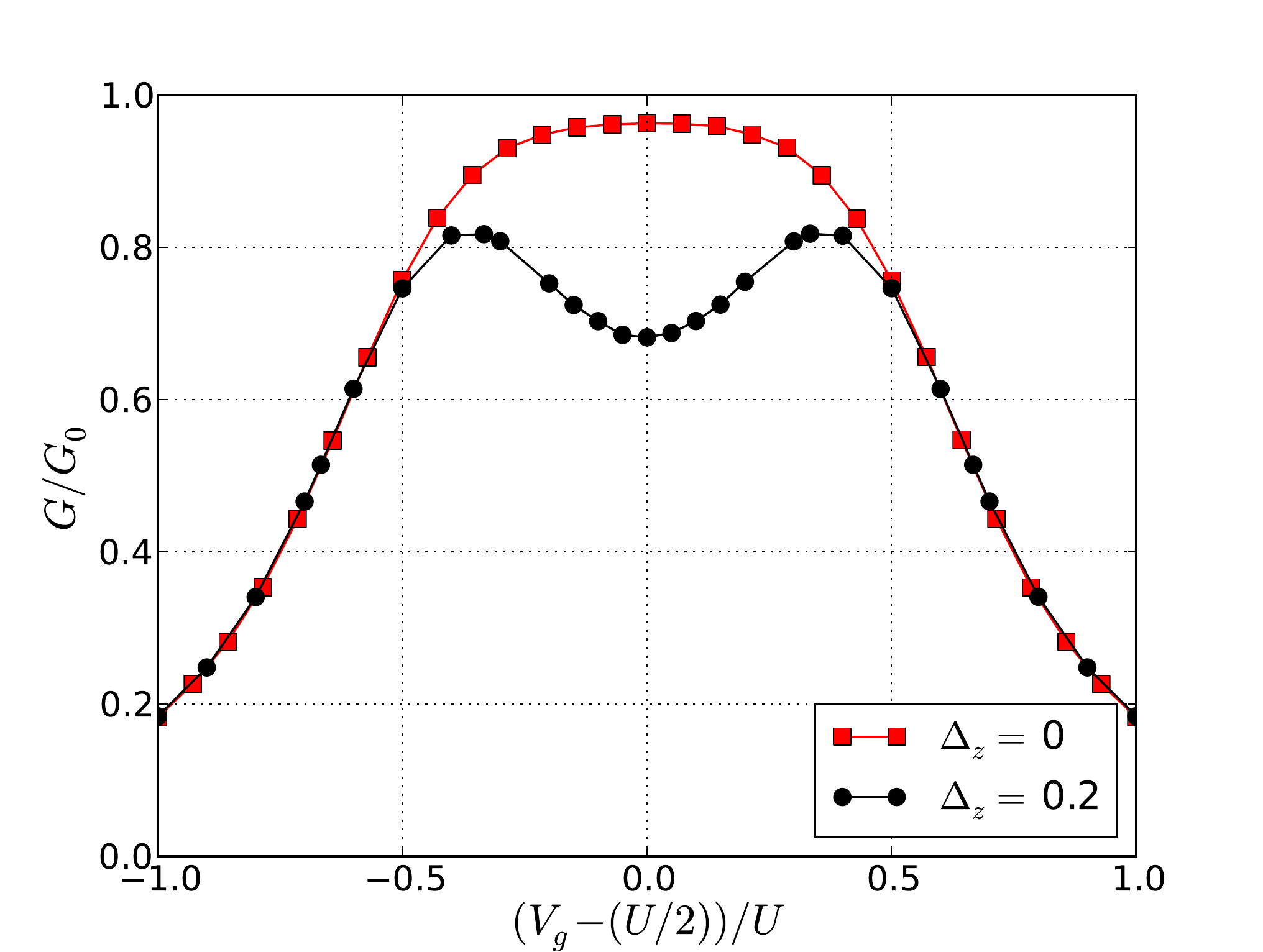}
\caption{Conductance as a function of gate voltage for finite ballistic wires (without mismatching, $t_{C}=t$) and two different 
values of the magnetic field. Other parameters are $V_g=0$ and  $T=0.01$. }\label{figb}
\end{figure}

In Fig.~\ref{figb} we illustrate the effect of the magnetic field on the behavior of the conductance as a function of the gate voltage 
at low temperature $T \ll T_K^0$. We see that the magnetic field tends to suppress the Kondo effect. In particular, the plateau of 
$G \sim G^0$ is suppressed giving rise to a valley between the two Coulomb blockade features.   

For $V_g=0$, and assuming a fixed temperature $T \ll T_K^0$, the conductance is a decreasing function of the magnetic field.
The effect of the magnetic field is to polarize the spin of the electrons at the dot. This effect conspires against the formation of 
the spin singlet, which is the origin of the Kondo resonance. As the magnetic field increases, the Kondo resonance melts down. In order 
to quantify this mechanism, we define the critical magnetic field $B_c$ as that for which $G(T,B_c)= G(T,0)/2$.  

\begin{figure}[htb]
\centering 
\includegraphics[clip,width=1\linewidth]{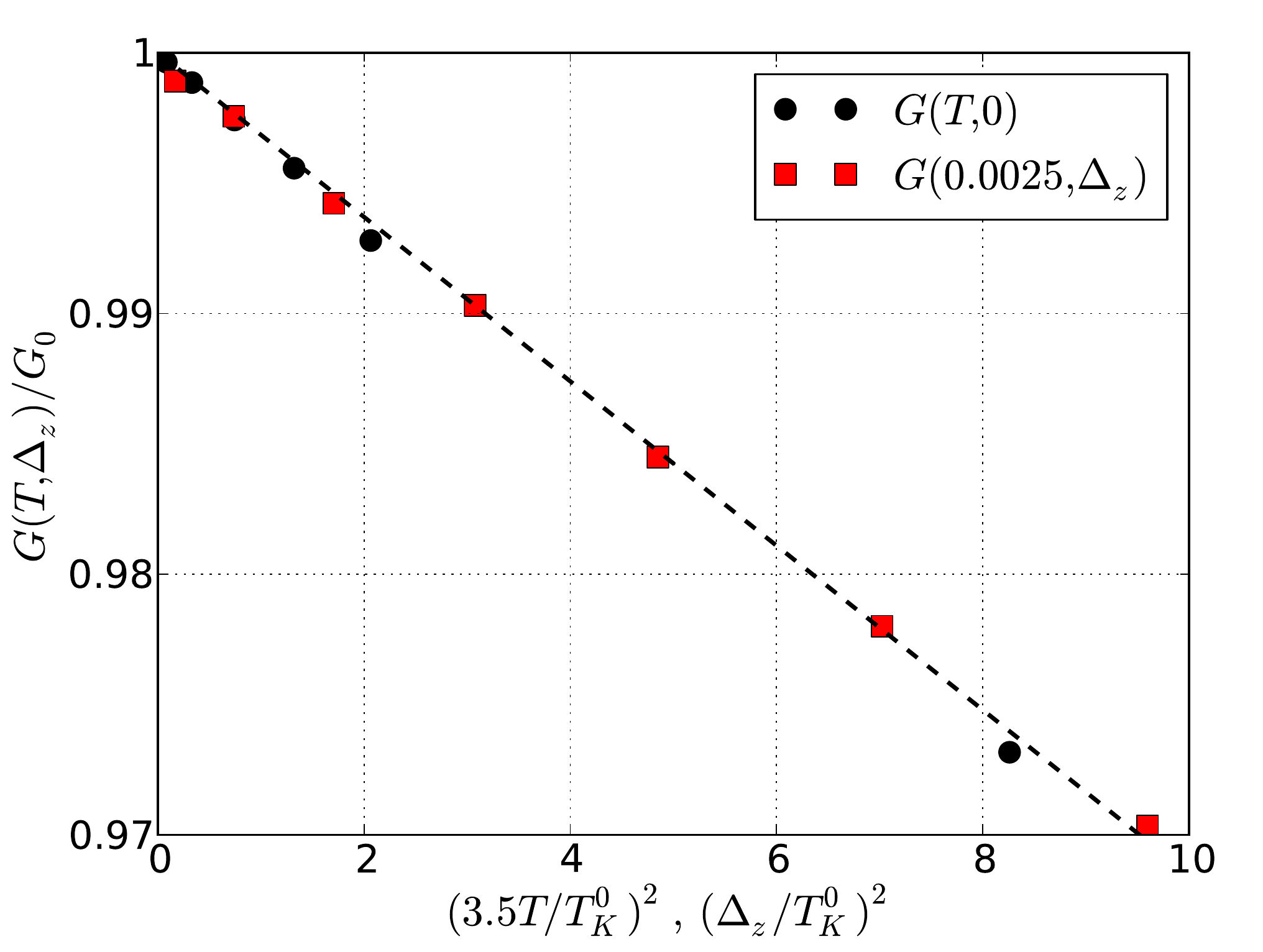}
\caption{Conductance as a function of the temperature squared $T^2$ and as a function of the Zeeman splitting squared 
$\Delta_z^2$ for a dot  connected to ballistic wires. The symbols correspond to quantum Monte Carlo data, while the line 
correspond to a quadratic fit with coefficients $c_T$ and $c_B$ as in Eq.~(\ref{gbt}).} \label{gscal}
\end{figure}

The above figures show that, without finite-size wires, the magnetic field plays a role in the behavior of the conductance which is 
similar to that played by the temperature. 
In fact, in Ref.~\onlinecite{smir-grif} a calculation based on a slave-boson treatment of the Kondo impurity and Fermi liquid theory 
shows the following universal law for the behavior of the ``zero-bias'' conductance as a function of temperature and magnetic field
\be \label{gbt}
\frac{G(T, \Delta_z)}{G^0} = \left[ 1- c_T (\frac{T}{T_K^0})^2 - c_B (\frac{\Delta_z}{T_K^0} )^2\right], \;\;\; |B|/T_K^0 < 1,
\ee 
where $c_T$ and $c_B$ are coefficients that depend on the Kondo temperature $T_K^0$ as well as on the hybridization with the wires 
$t^2 \rho_0(\mu)$, which satisfy the universal ratio 
\be \label{urat}
\frac{c_T}{c_B} \sim \pi^2.
\ee 

In Fig.~\ref{gscal} we show that the results of QMC are consistent with the scaling behavior predicted by Eq. (\ref{gbt}). In fact, 
we see that for low temperature and low magnetic field the
QMC data can be fitted with a quadratic function of $T$ and $B$. With a rescaling of the temperature by a factor close to $\pi$, both 
sets of data in the figure are fitted by the same quadratic function, which is in agreement with the ratio~(\ref{urat}).

\begin{figure}[htb]
\centering 
\includegraphics[clip,width=1\linewidth]{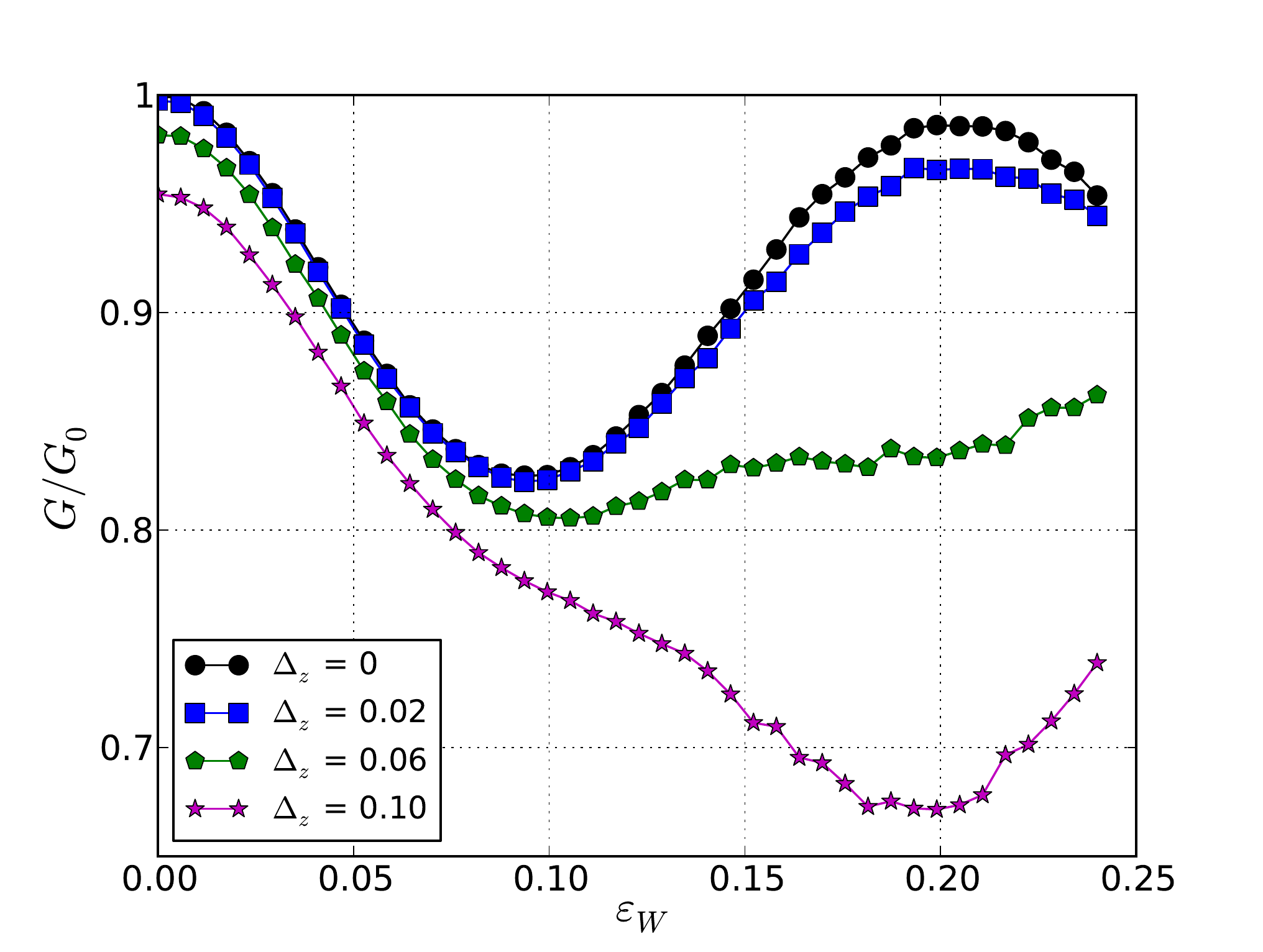}
\caption{Conductance as a function of  the local energy $\ve_W$ of the wires for 
 finite wires with mismatching $t_{C}=0.8t$ and, different values of the magnetic field.
Other parameters are $V_g=0$ and  $T=0.0025$. }\label{figbw}
\end{figure}

Equation~(\ref{gbt}) and the behavior of Fig.~\ref{figb} suggest that the magnetic field plays a similar role as the temperature 
regarding the behavior of the conductance. If we keep the temperature low $T \ll T_K^0$, we can infer that the ratio $\Delta/B_c$ sets 
the scale for the mesoscopic effects to play a role in the behavior of the conductance. 
In particular, in analogy to the behavior for $B=0$ and finite temperature analyzed in Sec.~\ref{sin_campo}, we can expect that for 
$\Delta \ll B_c$, the conductance is insensitive to the details of the wires, like the value of the local energy of the wires $\ve_W$ 
relative to $V_g$. However, for $\Delta \gtrsim B_c$, we expect to observe mesoscopic features in the behavior of the conductance as a 
function of the magnetic field.
In Fig.~\ref{figbw} we show the conductance as a function of the local energy of the wires $\ve_W$ for different values of $B$. 
For $B \ll B_c$ we observe a periodicity of $\Delta/2$. This is consistent with a perfect transmission when a level of the wires is 
resonant as well as when the Fermi energy is exactly off-resonant. This is precisely the behavior we discussed for $B=0$ 
in relation to Fig.~\ref{figos0} when  the temperature is $T \ll T_K^0$.
As the magnetic field increases above the critical value $B_c$ and the Kondo resonance disappears, the conductance changes its 
periodicity to that determined by the level spacing $\Delta$ of the wires. This change in the periodicity is similar to the one taking 
place as a function of the temperature shown in Fig.~\ref{figos0}.

\begin{figure}[htb]
\centering
\includegraphics[clip,width=1\linewidth]{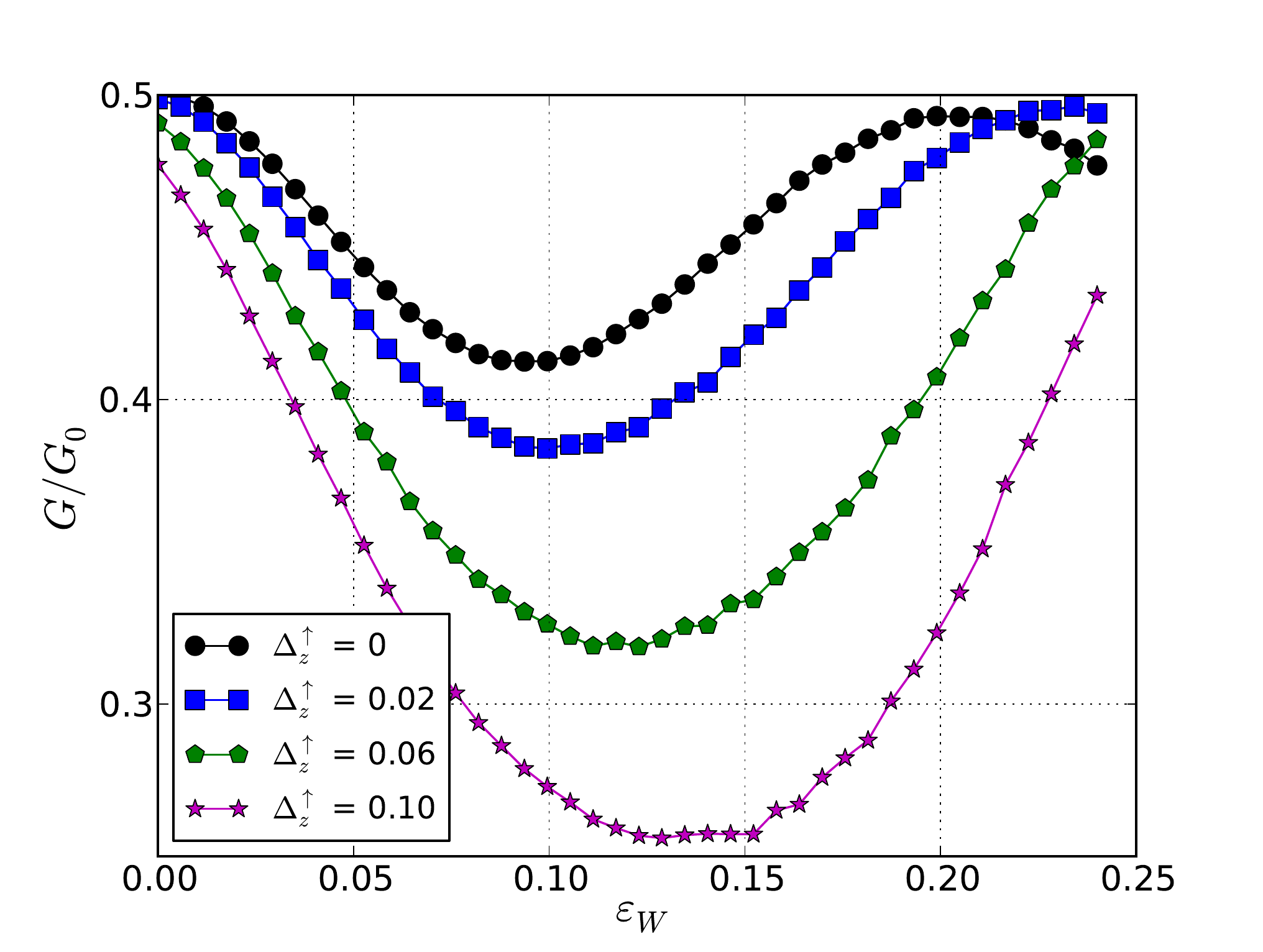}
\includegraphics[clip,width=1\linewidth]{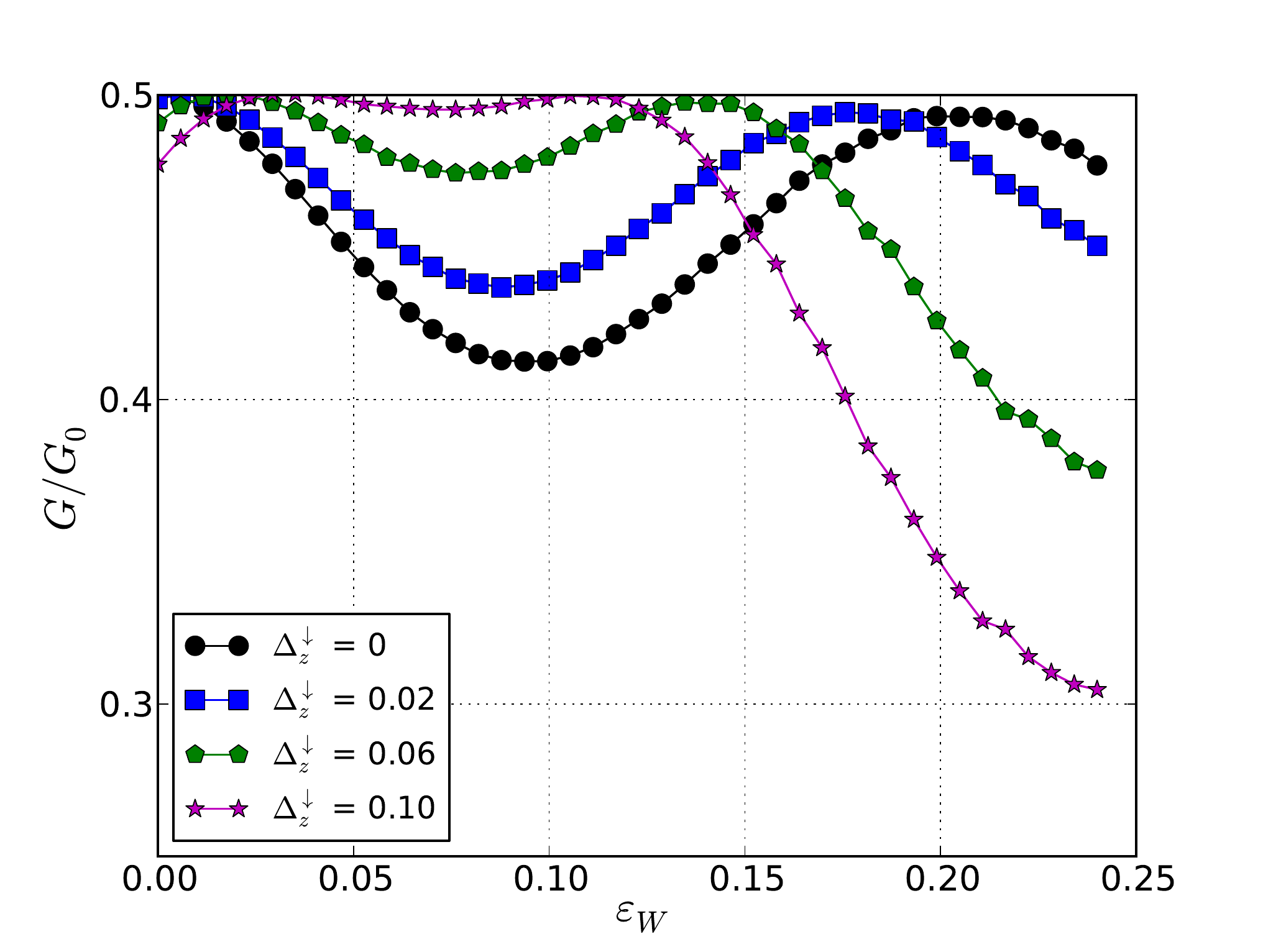}
\caption{Conductance for each spin species as a function of the local energy $\ve_W$ of the wires for 
 finite wires with mismatching $t_{C}=0.8t$ and, different values of the magnetic field.
Other parameters are $V_g=0$ and  $T=0.0025$.} \label{figbudw}
\end{figure}

In Fig.~\ref{figbudw} we show the conductance for each spin species corresponding to the same parameters of Fig.~\ref{figbw}. These plots satisfy
$G_{\uparrow}+ G_{\downarrow} = G$, where $G$ is the one shown in Fig.~\ref{figbw}.  For $B=0$, the Kondo effect is robust and both contributions to the conductance show the same pattern of oscillations with a period $\Delta/2$, and maxima when the levels of the wires are resonant or off-resonant with
the Kondo peak. As $B$ is switched on, a splitting of the Kondo peak is expected to take place. These changes reflect the changes in the behavior of the local density of states for the up and down spins at the dot close to the Fermi energy, $\rho_{\uparrow, \downarrow}(\omega), \; \; \omega \sim \mu$. For the largest value of $B$ shown in the Fig.~\ref{figbudw} (see plots in stars), the Kondo effect is expected to be broken and the emerging picture for the behavior of the conductance is the following. For $\varepsilon_W=0$, the density of states of the wires has a resonant peak at the Fermi energy, while the resonant peaks of $\rho_{\uparrow, \downarrow}(\omega)$ are shifted to $\omega= \pm \Delta_z $, respectively.  The effect of increasing $\varepsilon_W$ is a downwards rigid shift  of the density of states of the wires. For $\varepsilon_W$ increasing from zero,
 there is a resonant peak of the density of states of the wires, which was originally centered at $\omega=0$ and tends to get aligned with the resonant peak of $\rho_{\downarrow}(\omega)$ at the same time that it further departs from the one of $\rho_{\uparrow}(\omega)$. The result is an increasing (decreasing) conductance for the down (up) spins. For even larger 
$\varepsilon_W \sim \Delta/2$ the density of states of the wires has a valley at the Fermi energy and, as the Kondo effect is not active, the conductance decreases for both spins. For larger
values of $\varepsilon_W$ within the range $\Delta/2< \varepsilon_W < \Delta$, a resonant peak of the wires which was originally at $\omega=\Delta$ is shifted downwards and tends to be aligned with the peak at $\omega= \Delta_z$ of $\rho_{\uparrow}(\omega)$. Hence, the conductance of $\uparrow$ spins increases, while the one for $\downarrow$ spins
continues the decreasing behavior.

\section{Summary and Conclusions}
We have studied the mesoscopic features in the conductance of a quantum dot with a magnetic field, modeled by an Hubbard-Anderson impurity model connected to finite one-dimensional wires.
We have focused on the linear regime in the bias voltage and evaluated the conductance by recourse to a quantum Monte Carlo method. In order to benchmark this numerical 
technique, we have first focused on a vanishing magnetic field and identified the change in the period of the mesoscopic conductance oscillations as the temperature grows above the
Kondo temperature. 

We have then studied the effect of the magnetic field. In the absence of mismatching between the wires and the reservoirs, we have recovered universal relations in the behavior of the conductance as a function of the temperature and the magnetic field. These relations indicate that the magnetic field and the temperature play a similar role in the behavior of the conductance. In particular, both quantities have the effect of destroying the Kondo singlet, thus tending to decrease the conductance. Furthermore, the departure from the perfect conductance quantum is quadratic both in the temperature and the magnetic field. It is, thus, not surprising that the magnetic field may introduce similar effects in the behavior of the
mesoscopic features in the presence of finite-size wires. In fact, the conductance displays oscillations as a function of a gate voltage applied at the wires. As the magnetic field overcomes
a critical value, for which the Kondo effect is sufficiently weak, the period of such oscillations change to twice its original value as happens at zero magnetic field and the temperature 
increases over the Kondo temperature. 

The change in the behavior of these mesoscopic oscillations of the conductance as a function of the temperature, have been proposed as a way to detect the Kondo 
screening length, relative to the length of the  wires.\cite{sim-af} The fact that varying a magnetic field instead of the temperature has a similar effect is interesting from
the experimental point of view. Notice that the Kondo temperature in quantum dots is of the order of some few mK, hence it is difficult to control changes of the temperature within
such a small scale. It is, however, much easier to control changes in the magnetic field within the range set by the critical field for the Kondo effect, which is of the order of 
100 mT.\cite{18}

\section{Acknowledgements}
This work is supported by CONICET, MINCyT and UBACYT, Argentina.

\appendix

\section{Calculation of the non-interacting retarded Green functions} \label{nonint}
An  exact procedure is to numerically evaluate the Matsubara and retarded Green functions $g_{\alpha}(i \omega_n)$ and $g^R_{\alpha}(\omega)$
is the recursive solution of the Dyson equation $g_{\alpha}(i \omega_n)= g_{1,\alpha}(i \omega_n)$, with
\begin{equation}
g_{l,\alpha} (i \omega_n)^{-1}=i \omega_n -\varepsilon_{W} -t^2 g_{l+1,\alpha}(i \omega_n),
\end{equation}
for $1 \leq l < N_{\alpha}$ and 
\begin{equation}
g_{N_{\alpha},\alpha}(i \omega_n)^{-1}= i\omega_n-\varepsilon_{W}- t_{C}^2 g^0_{\alpha}(i \omega_n),
\end{equation}
where  $g^0_{\alpha}(i \omega_n)$ is the Matsubara Green function corresponding to a semi-infinite tight-binding chain with hopping $t$. 
The corresponding expressions for the retarded functions can be easily obtained by substituting $i \omega_n \rightarrow \omega + i \eta$.



\end{document}